\documentclass[conference]{IEEEtran}
\IEEEoverridecommandlockouts

\usepackage{cite}
\usepackage{amsmath,amssymb,amsfonts}
\usepackage{algorithmic}
\usepackage{graphicx}
\usepackage{textcomp}
\usepackage{xcolor}
\usepackage[hidelinks]{hyperref}

\def\BibTeX{{\rm B\kern-.05em{\sc i\kern-.025em b}\kern-.08em
    T\kern-.1667em\lower.7ex\hbox{E}\kern-.125emX}}

\makeatletter
\newcommand{\linebreakand}{%
  \end{@IEEEauthorhalign}
  \hfill\mbox{}\par
  \mbox{}\hfill\begin{@IEEEauthorhalign}
}
\makeatother

\begin{document}

\title{WIP: Large Language Model-Enhanced Smart Tutor for Undergraduate Circuit Analysis
}

\author{
\IEEEauthorblockN{Liangliang Chen, Huiru Xie, Jacqueline Rohde, Ying Zhang$^{\dagger}$\thanks{\textcopyright 2025 IEEE. Accepted to 2025 Frontiers in Education (FIE) Conference. Personal use of this material is permitted. Permission from IEEE must be obtained for all other uses, in any current or future media, including reprinting/republishing this material for advertising or promotional purposes, creating new collective works, for resale or redistribution to servers or lists, or reuse of any copyrighted component of this work in other works.}}
\IEEEauthorblockA{\textit{School of Electrical and Computer Engineering}, \textit{Georgia Institute of Technology}, Atlanta, USA \\
$^{\dagger}$Corresponding Author. E-mail: \texttt{yzhang@gatech.edu}}}

\maketitle

\begin{abstract}
This research-to-practice work-in-progress (WIP) paper presents an AI-enabled smart tutor designed to provide homework assessment and feedback for students in an undergraduate circuit analysis course. We detail the tutor’s design philosophy and core components, including open-ended question answering and homework feedback generation. The prompts are carefully crafted to optimize responses across different problems. The smart tutor was deployed on the Microsoft Azure platform and is currently in use in an undergraduate circuit analysis course at the School of Electrical and Computer Engineering in a large, public, research-intensive institution in the Southeastern United States. Beyond offering personalized instruction and feedback, the tutor collects student interaction data, which is summarized and shared with the course instructor. To evaluate its effectiveness, we collected student feedback, with 90.9\% of responses indicating satisfaction with the tutor. Additionally, we analyze a subset of collected data on preliminary circuit analysis topics to assess tutor usage frequency for each problem and identify frequently asked questions. These insights help instructors gain real-time awareness of student difficulties, enabling more targeted classroom instruction. In future work, we will release a full analysis once the complete dataset is available after the Spring 2025 semester. We also explore the potential applications of this smart tutor across a broader range of engineering disciplines by developing improved prompts, diagram-recognition methods, and database management strategies, which remain ongoing areas of research.
\end{abstract}


\section{Introduction}

Large language models (LLMs) have been widely used to support student learning across various educational fields, such as programming \cite{savelka2023thrilled}, robotics \cite{chen2024rlingua}, medical education \cite{kung2023performance}, and legal studies \cite{guha2023legalbench}. However, their potential applications in engineering disciplines remain underexplored. Leveraging LLMs in engineering education enables automated, context-aware feedback on problem-solving processes, enhancing student learning with personalized guidance and real-time assessments. LLMs also facilitate interactive tutoring and scalable grading, reducing instructor workload while maintaining educational quality.

Developing automated educational tools that provide adaptive instruction and guidance to students has been a long-standing goal \cite{petrina2004sidney}. Before the era of LLMs, these tools were primarily built using rule-based methods. For example, the Andes system \cite{schulze2000andes}, an intelligent tutor for classical physics, leveraged goal rules and physics-knowledge rules to generate necessary equations for problem-solving, enabling it to provide immediate feedback and relevant hints to students. Skromme \textit{et al.} \cite{skromme2013computer, skromme2014expansion} developed a tutoring system for linear circuit analysis that could generate new circuit problems with randomized topologies and element values. By employing rule-based strategies, these systems ensured error-free solutions; however, students were restricted to following predefined rules—such as filling in blanks—rather than freely expressing their solutions or asking personalized, open-ended questions. Some studies have incorporated deep learning methods to model students' knowledge during the learning process \cite{piech2015deep} and provide exercise problem recommendations \cite{ai2019concept}. However, these approaches have not extended personalization to allow students to ask open-ended, individualized questions and receive tailored support. The advent of LLMs presents new opportunities for building smart tutors that closely resemble real-person instructors across various disciplines \cite{achiam2023gpt}. However, most LLM-based educational tools have been developed in fields where LLMs perform well, such as programming \cite{savelka2023thrilled} and medical education \cite{kung2023performance}, while comparatively fewer efforts have focused on the broader field of engineering education. 

\begin{figure*}[!t]
\centering
\includegraphics[width=0.875\textwidth]{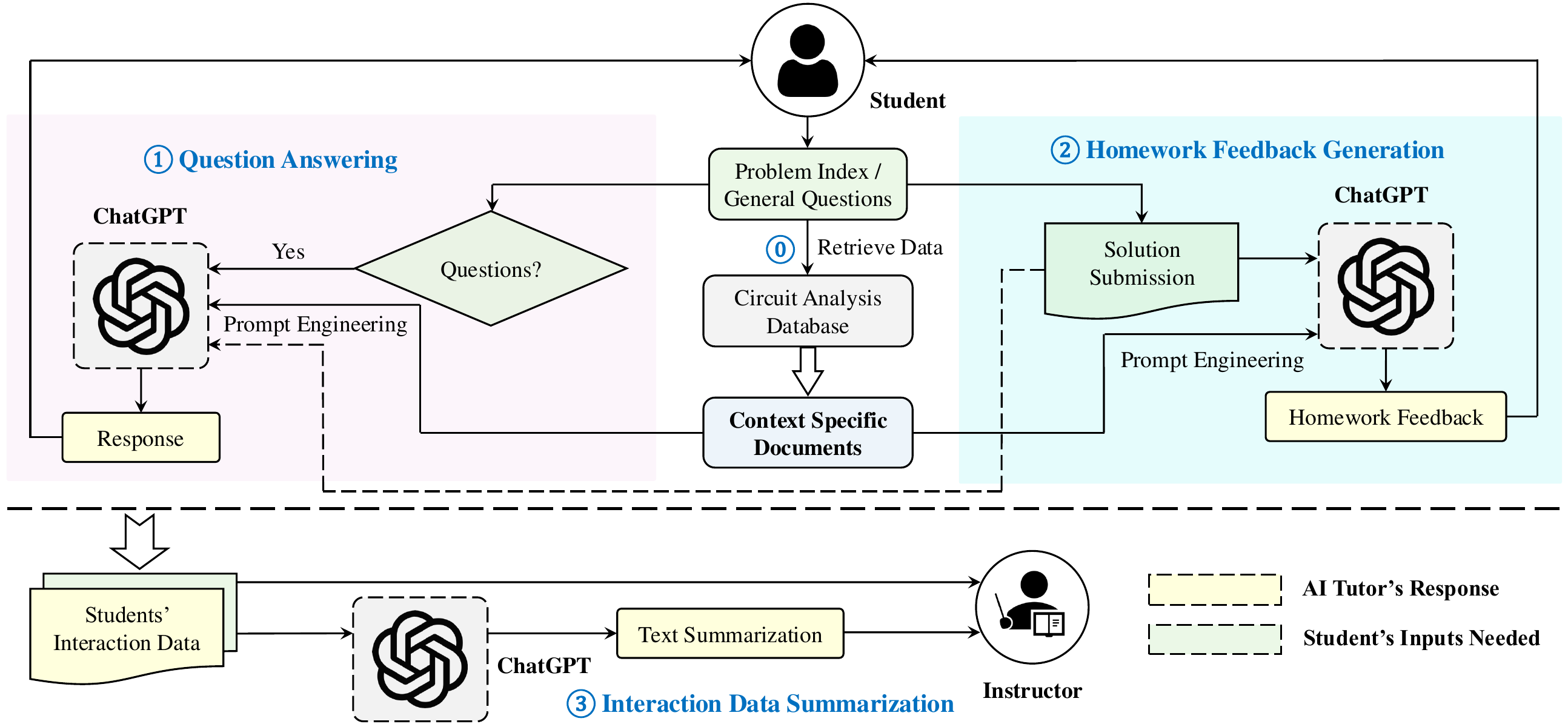}
\caption{Framework of the Smart Tutor: \raisebox{.5pt}{\textcircled{\raisebox{-.9pt} {0}}} Upon receiving a problem index or a general question from the student, the system retrieves the corresponding context-specific documents from the circuit analysis database. \raisebox{.5pt}{\textcircled{\raisebox{-.9pt} {1}}} Students can ask open-ended questions at any time during the homework completion process. \raisebox{.5pt}{\textcircled{\raisebox{-.9pt} {2}}} Students may submit their homework solutions to receive feedback. The submitted solution is also incorporated into the prompts of the open-ended question-answering LLMs to support post-submission queries. \raisebox{.5pt}{\textcircled{\raisebox{-.9pt} {3}}} The interaction data between students and the smart tutor is summarized and reported to the instructor.}
\label{framework}
\end{figure*}

The main challenges of integrating LLMs into engineering education include their inability to recognize and interpret scientific and engineering diagrams \cite{hou2024vision}, limited mathematical capabilities \cite{ahn2024large}, and the prevalence of hallucinations \cite{huang2025survey}. Importantly, reliability is a key consideration when designing automated educational support tools, as incorrect responses can mislead students and negatively impact their learning, particularly for those who are struggling \cite{rohde2024predictors}. In this paper, we investigate how a smart tutor for an undergraduate circuit analysis course can be developed with the assistance of LLMs. This smart tutor demonstrates high reliability in providing homework support at different stages of learning.

The contributions of this paper are summarized as follows:

\begin{itemize}
\item [i)] We present a design framework for an LLM-enhanced smart tutor. Currently focused on an undergraduate circuit analysis course, the smart tutor provides real-time homework guidance, question answering, and solution feedback by leveraging appropriate offline documents and prompts. The interaction data between students and the smart tutor are stored in a database. By analyzing this data from both student and problem perspectives, instructors can be notified of frequently asked questions, common areas of difficulty, and personalized weaknesses of individual students. This enables instructors to provide more targeted guidance and support based on insights generated by the smart tutor.

\item [ii)] We then describe the deployment of the smart tutor using the Microsoft Azure platform in an undergraduate circuit analysis course during the Spring 2025 semester in the School of Electrical and Computer Engineering at a large, public, research-intensive institution in the Southeastern United States. Preliminary data from student feedback and interaction data indicate that the smart tutor is helpful in supporting students' learning and homework completion. We discuss directions for ongoing work.
\end{itemize}

\section{Smart Tutor Design}
\label{S2}
This section introduces the design philosophy and basic components of the smart tutor. Fig. \ref{framework} shows the framework of the smart tutor designed in this paper. 

\subsection{Design Philosophy}
\label{S21}
Ref. \cite{chen2025benchmarkinglargelanguagemodels} demonstrated the strong performance of GPT-4o\footnote{See https://cdn.openai.com/gpt-4o-system-card.pdf.}—when grounded with relevant context, such as reference solutions—in providing homework assessments based on various metrics for undergraduate circuit analysis. The authors also identified several limitations of LLMs in this domain, including difficulties in recognizing and interpreting circuit diagrams, processing both handwritten and typed homework submissions, and executing mathematical computations. The findings in \cite{chen2025benchmarkinglargelanguagemodels} suggest that baseline LLMs must be enhanced to provide reliable homework support for students in circuit analysis courses. To address this need, we develop an offline repository of essential documents that equips LLMs with sufficient problem- and topic-specific context to generate accurate and relevant responses. A detailed description of this offline document database is provided in Section \ref{S22}.

Students' needs vary depending on factors such as individual learning ability, background knowledge, and the stage of homework completion. To address these variations, the smart tutor pipeline comprises two main components: i) real-time open-ended question answering and ii) homework feedback generation, as illustrated in Fig. \ref{framework}. These components are discussed in detail in Sections \ref{S23} and \ref{S24}, respectively.

\subsection{Circuit Analysis Context Specific Database}
\label{S22}
State-of-the-art LLMs currently lack the capability to reliably recognize and interpret circuit diagrams \cite{chen2025benchmarkinglargelanguagemodels}. Consequently, they are unable to independently solve circuit analysis problems or provide feedback based solely on textual descriptions, as such problems typically rely heavily on visual circuit representations. Moreover, LLMs may also exhibit hallucinations when responding to general questions posed by students. To mitigate these limitations, an offline database can be prepared containing contextual information—specific to various circuit problems or topics—in text format. This dataset can then be retrieved and incorporated into the LLMs' prompts, enabling the generation of accurate and relevant responses.

In general, this database can take one of two formats, depending on the context and data availability:

\begin{itemize} 
\item [i)] \textit{Structured Database:} This type of database stores structured or tabular data, allowing for exact matching. When using this approach, problem- or topic-specific data must be precompiled by the course instructor. The advantage of this format is that exact matching ensures the relevance of the retrieved data. However, the downside is that preparing such a database can be labor intensive.
\item [ii)] \textit{Vector Database:} Unlike a structured database, a vector database stores numerical data embeddings, which are retrieved using similarity-based search \cite{pan2024survey}. This approach is less labor intensive in terms of data labeling and preparation. However, the quality of the retrieved data depends on several factors, such as the embedding method, similarity metric, and approximate nearest neighbor algorithm.
\end{itemize}

\subsection{Homework Support via Open-Ended Question Answering}
\label{S23}
During the process of solving circuit problems, students may struggle with identifying the correct approach or have general concept-related questions. To support them, a chat window can be integrated into the smart tutor, offering open-ended question-answering support as needed. The level of assistance can be adapted to individual students’ needs or estimated knowledge levels—for example, by providing either high-level problem-solving strategies or detailed, step-by-step explanations.

Students may pose questions at any point during their interaction with the system. Before solution submission, the retrieved subset from the context-specific database can be appended to the LLMs' prompts to generate context-aware responses. After submission, the student's solution is also included in the prompts, enabling personalized feedback on their work in addition to support on general concepts. To further enhance response quality—particularly for concept-based questions—the retrieval-augmented generation (RAG) framework \cite{gao2023retrieval} can be employed, utilizing similarity-based information matching from a vector database. Additionally, LLMs may be prompted to reference lecture notes when responding to queries. However, they are explicitly instructed not to disclose reference solutions to students.

\subsection{Homework Feedback Generation}
\label{S24}
Once students have completed their solutions, they can submit them for feedback. LLMs provide structured homework feedback in multiple rounds, evaluating different aspects of the solution. Inspired by \cite{chen2025benchmarkinglargelanguagemodels}, our implementation assesses solutions using the following four metrics:

\begin{itemize} 
\item [i)] \textit{Final Answer and Arithmetic Accuracy:} Evaluates whether the student's final answer is correct and checks for arithmetic errors throughout the solution. If an error is detected, the LLM identifies and explains it.

\item [ii)] \textit{Completeness:} Determines whether the student has fully answered all sub-questions in the problem.

\item [iii)] \textit{Method:} Assesses whether the student has applied the correct problem-solving method, regardless of arithmetic or typographical errors.

\item [iv)] \textit{Units:} Many students omit or use incorrect units for variables. The LLM verifies unit correctness and provides feedback accordingly.
\end{itemize}

By providing feedback from multiple perspectives, LLMs generate more precise and reliable evaluations compared to a single round of general feedback \cite{chen2025benchmarkinglargelanguagemodels}. However, presenting all rounds in full may be overwhelming, particularly for students with correct solutions. To address this, the LLM can be prompted to summarize its feedback across these four aspects. By default, the smart tutor displays the summarized feedback, with the detailed breakdown available upon request.

\subsection{Interaction Data Summarization}
\label{S26}
The smart tutor's components in Sections \ref{S23} and \ref{S24} are designed to facilitate student interactions, providing real-time homework and learning support. The interaction data is stored and can be summarized using LLMs before being shared with instructors. These summaries can be generated at both the problem and student levels. At the problem level, common issues and frequently asked questions across different students can be identified, helping instructors clarify relevant circuit concepts and methods in class. At the student level, personalized learning progress and needs can be analyzed, enabling tailored support in subsequent learning processes \cite{sun2025data}. 

\section{Preliminary Implementation and Results}
Section \ref{S2} describes the basic components of the designed smart tutor. We deployed the tutor on the Microsoft Azure platform\footnote{See https://azure.microsoft.com.} and offered it as an optional resource during the Spring 2025 semester to students in an undergraduate circuit analysis course at the School of Electrical and Computer Engineering in a large, public, research-intensive institution in the Southeastern United States. The data collection and analysis procedures were reviewed and approved by the university's Institutional Review Board (IRB H23382). Of the 53 students in the class, 51 voluntarily registered for the smart tutor. We present and analyze students' interaction data from Homework \#1, which consists of nine problems covering electric circuit variables, circuit elements, and resistive circuits, corresponding to Chapters 1–3 in \cite{svoboda2013introduction}. 

Our deployment uses GPT-4o as the LLM. We precompile a structured database that can be retrieved to provide supporting documents for the LLM. In the pre-submission stage, students can choose from different levels of assistance based on their needs, including general method hints, step-by-step instructions, and other open-ended questions. When requesting homework feedback, students are required to format their solutions as text, with equations written in either LaTeX or plain text.

\subsection{Feedback from Students}

\begin{figure}
	\centering
	\includegraphics[width=0.35\textwidth]{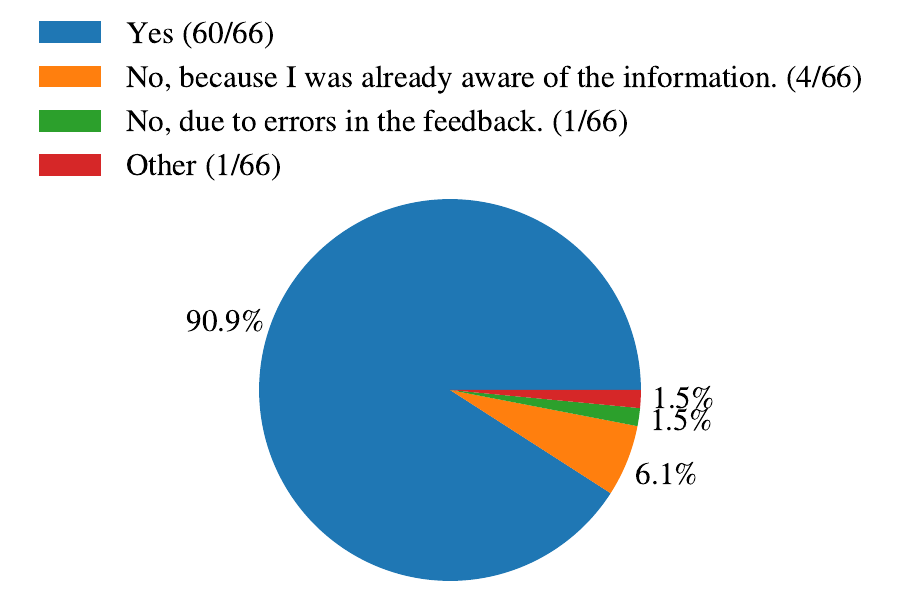} 
	\caption{Proportions of different types of student responses to the question, ``\textit{Do you find the homework feedback useful?}"} 
	\label{Fig2} 
\end{figure}

After each problem session, we ask students, ``\textit{Do you find the homework feedback useful?}" to gather their feedback on their experience. During the week of Homework \#1 completion, we collected 66 responses from students. Fig. \ref{Fig2} shows the proportions of different types of responses.

As shown in Fig. \ref{Fig2}, 90.9\% of responses indicate that the smart tutor's feedback was helpful, while 6.1\% state that the students were already aware of the information before receiving the feedback. This suggests that most students had a positive experience using the smart tutor. Only one response out of 66 reported errors in the feedback. Additionally, one student noted, ``\textit{My answer was the same as the one shown as correct by the AI tutor; however, the tutor said it was incorrect,}" which we categorize as ``Other". These errors are due to GPT-4o’s hallucinations, which we are actively working to mitigate. 

\subsection{Interaction Data Analyses}

As stated in Section \ref{S26}, interaction data between students and the tutor can be stored and analyzed to provide insights for course instructors, enabling them to offer timely, tailored support in subsequent class sessions. One format of the summarized data includes the number of students who sought help before and after homework submissions and requested feedback. Assuming that more students ask questions about more difficult problems, this type of summarization can help instructors identify which problems—and the related concepts or methods—need to be reiterated in class. 

\begin{figure}[!t]
	\centering
	\includegraphics[width=0.35\textwidth]{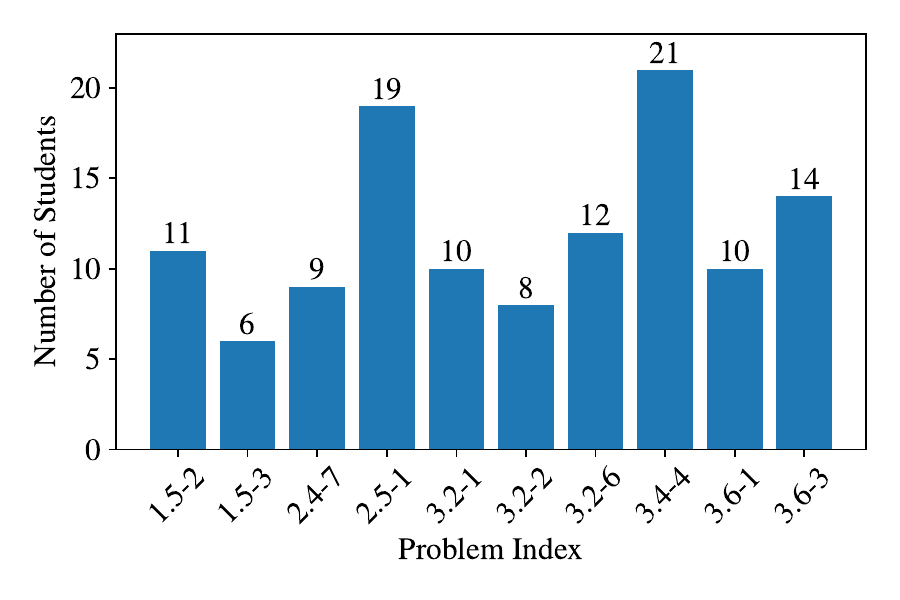}
	\caption{Number of students asking open-ended questions before homework submission}
	\label{Fig3}
\end{figure}

\begin{figure}[!t]
	\centering
	\includegraphics[width=0.35\textwidth]{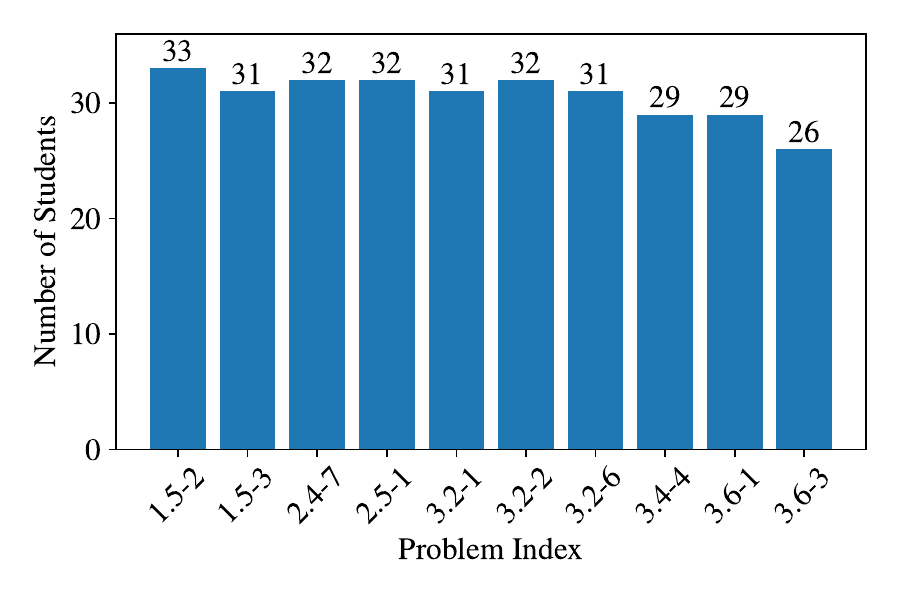}
	\caption{Number of students requesting homework feedback}
	\label{Fig4}
\end{figure}

\begin{figure}[!t]
	\centering
	\includegraphics[width=0.35\textwidth]{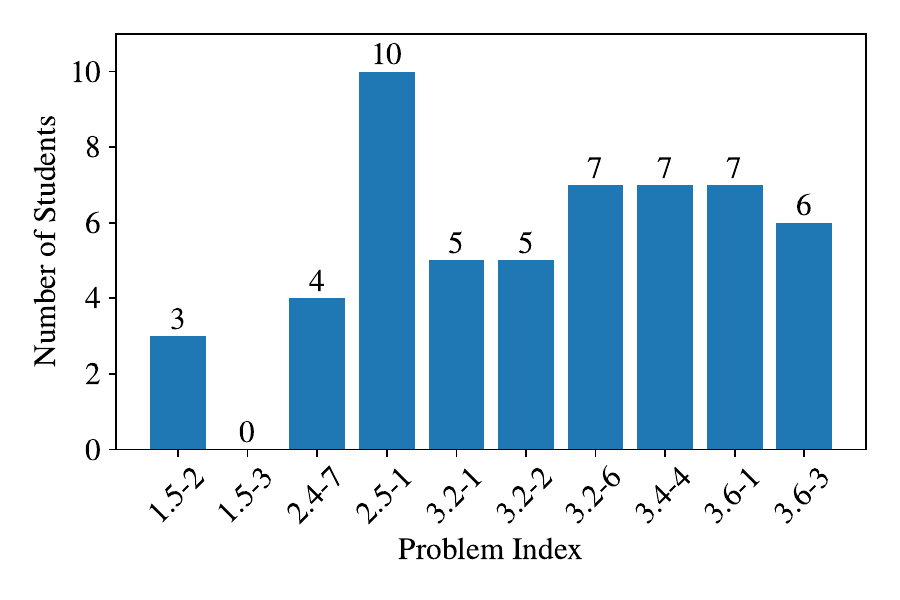}
	\caption{Number of students asking open-ended questions after homework submission}
	\label{Fig5}
\end{figure}

Figs. \ref{Fig3}–\ref{Fig5} present the numbers of students who asked open-ended questions before and after homework submission, as well as those who requested homework feedback, for various problems in Homework \#1. While the number of students requesting feedback is relatively consistent across different problems, the number of open-ended questions varies significantly between the pre- and post-submission phases. For pre-submission questions, a greater number of students sought support for problems 2.5-1 and 3.4-4, whereas post-submission questions were more about problem 2.5-1. These patterns suggest that instructors should closely evaluate the difficulty levels of problems 2.5-1 and 3.4-4, along with the associated concepts or methods, in order to identify student difficulties and address them promptly during class sessions.

Apart from analyzing the number of students seeking support at different stages, we can also identify frequently asked questions from the interaction data. For Homework \#1, the following questions were most commonly asked by students:

\begin{itemize} \item \textit{Questions Asked Before Solution Submission}
	 \begin{itemize} 
	 	\item [--] What is current/voltage division, and how is it used?
		\item [--] How can I use KCL to find the current $i$?
		\item [--] I am not sure how to proceed with this question.
	\end{itemize} 
	\item \textit{Questions Asked After Solution Submission} 
	\begin{itemize} 
		\item [--] What is the passive sign convention, and how does it apply to this problem?
		\item [--] How do I apply the current division principle to this problem?
		\item [--] How can I calculate power without knowing the current?
	\end{itemize} 
\end{itemize}

Instructors can then provide further explanations for the summarized questions above. 

\section{Discussions on Ongoing Work}
As a work-in-progress (WIP) paper, we present the framework of the smart tutor, along with our preliminary deployment and results. The high percentage of positive student feedback, along with an analysis of student-tutor interactions, demonstrates the practical value of our system.

As we continue to improve the smart tutor, our ongoing work focuses on the following areas:

\begin{itemize}
\item [i)] Reducing LLM hallucinations, which occasionally occur in areas such as rounding error identification, unit conversions, and arithmetic deductions.

\item [ii)] Developing more efficient database management methods to minimize the labor required for data preparation.

\item [iii)] Enhancing the tutor’s ability to recognize and interpret circuit diagrams, incorporating extracted diagram information into the LLM's question-answering and homework feedback generation processes.

\item [iv)] Expanding the framework to other engineering courses, such as advanced microelectronic circuits, and incorporating diverse tutoring formats, such as practice problem recommendations.
\end{itemize}

\section{Conclusions}
This paper presented an AI-enabled smart tutor designed to support homework assessment and feedback in an undergraduate circuit analysis course. By integrating problem-specific homework question-answering and automated feedback generation, the tutor enhances students’ learning experiences while providing instructors with valuable insights into common student challenges. Initial deployment results indicate strong student satisfaction, with 90.9\% of surveyed students expressing positive feedback on the tutor’s effectiveness.

Our preliminary analysis of student interactions highlights the tutor’s potential to improve learning outcomes by identifying frequently encountered difficulties and enabling more targeted instruction. As we continue refining the system, we will explore its applicability to broader engineering disciplines by enhancing prompt design, incorporating circuit diagram recognition, and optimizing database management for improved response accuracy. Through these advancements, we aim to further bridge the gap between AI-driven tutoring and effective engineering education, fostering a more interactive and personalized learning environment.

\section{Acknowledgments}
The authors appreciate the support provided by the School of Electrical and Computer Engineering and the College of Engineering at the institution where the study was conducted. The authors would also like to acknowledge the assistance of ChatGPT in polishing the language of this paper.

\bibliographystyle{IEEEtran}
\bibliography{IEEEfull.bib}

\end{document}